\documentstyle[11pt]{article}
\begin{document}
\begin{titlepage}
\begin{centering}
 
{\ }\vspace{2cm}
 
{\Large\bf Quantisation and Gauge Invariance\\}
\vspace{2cm}
Jan Govaerts\footnote{E-mail: {\tt govaerts@fynu.ucl.ac.be}}\\
\vspace{1.0cm}
{\em Institut de Physique Nucl\'eaire}\\
{\em Universit\'e catholique de Louvain}\\
{\em 2, Chemin du Cyclotron}\\
{\em B-1348 Louvain-la-Neuve, Belgium}\\
\vspace{2cm}
\begin{abstract}

\noindent
Recent developments concerning canonical quantisation and gauge
invariant quantum mechanical systems and quantum field
theories are briefly discussed. On the one hand, it is shown how 
diffeomorphic covariant representations of the Heisenberg algebra 
over curved manifolds of non trivial topology involve topology classes 
of flat U(1) bundles. On the other hand, through some examples, 
the recently proposed physical projector approach to the quantisation of 
general gauge invariant systems is shown to avoid the necessity of 
any gauge fixing---hence also avoiding the possibility of Gribov problems 
which usually ensue any gauge fixing procedure---, and is also capable 
to provide the adequate description of the physical content of gauge
invariant systems.

\end{abstract}

\vspace{15pt}
 
To appear in the Proceedings of the\\
1$^{\rm st}$ International Workshop on Contemporary Problems in Mathematical
Physics,\\
Institut de Math\'ematiques et Sciences Physiques (IMSP), Universit\'e
Nationale du B\'enin,\\
Cotonou, R\'epublique du B\'enin\\
October 31$^{\rm st}$ - November 5$^{\rm th}$, 1999

\end{centering} 

\vspace{100pt}

%\noindent hep-th/9910xxx\\
\noindent October 1999

\end{titlepage}

\section{Introduction}
\label{Sect1}

Two of the most fundamental tenets of modern physics at the turn of this 
century are, on the one hand, the general concepts of quantum physics,
and on the other hand, the general idea of local or gauge symmetries
as the governing principle for all interactions and their properties as well
as for their ultimate unification.

Nevertheless, the combination of these two general conceptual frameworks
is not straightforward, and is in fact beset by a series of difficult
issues. For example, a manifest realisation of a local symmetry requires
the introduction of gauge variant degrees of freedom which cannot be associated
to physical, {\em i.e.\/} gauge invariant observables. Maintaining such
degrees of freedom explicit through the quantisation procedure requires tools 
which in effect cancel the contributions of these non phy\-si\-cal 
configurations 
to physical observables. On the other hand, reducing these gauge variant 
degrees of freedom before quantisation often runs counter to a manifest 
realisation of other symmetries, such as global spacetime symmetries as for 
example Poincar\'e invariance, while more importantly still, such a reduction
implies a gauge invariant configuration space description whose topology 
and geometry is typically very intricate and difficult to circumvent
in general terms, thereby leading to the possibility of Gribov 
problems\cite{Gov1}. At the quantum level, the difficulty is then 
exacerbated still further by the fact that quantisation techniques for 
manifolds of non tri\-vial topology and geometry are much more involved to 
develop than for the ordinary and most familiar case of a flat euclidean space.

In this contribution, we wish to discuss two recent developments concerned
with such issues. On the one hand, the general classification of 
re\-pre\-sen\-ta\-tions of the Heisenberg algebra over configuration space 
manifolds of arbitrary to\-po\-lo\-gy and geometry has recently been 
considered\cite{Gov2}, leading to the possibility of inequivalent 
representations associated to all topologically distinct flat U(1) 
bundles which may be defined for a base
manifold whose first homotopy group is non trivial. 

On the other hand, a new approach to the quantisation 
of constrained systems, which include gauge invariant ones, has recently
been suggested\cite{Klauder1}. This approach is formulated solely
within Dirac's quantisation of such 
systems\cite{Gov1}, does not require any extension (such as in BRST 
quantisation\cite{Gov1}) nor any reduction (such as in Faddeev's reduced 
phase space quantisation\cite{Gov1}) of the initial formulation of the system, 
and uses in an essential way the physical projection operator onto gauge 
invariant quantum states only.
Since the latter operator is defined through integration over the manifold
of the group of gauge transformations, it lends itself to general group
theoretical methods for the evaluation of its matrix elements, which are
also the generating functions for the wave functions of gauge invariant
states. Some of the advantages of this new approach to quantisation have already
been illustrated in some simple examples, showing\cite{Gov3} among other 
important points, that this method being free of the necessity of any gauge 
fixing procedure is also necessarily free of any Gribov of any type\cite{Gov1},
a situation which is not achieved with any of the other quantisation methods.
Until now, the physical projector approach to gauge invariant systems
has been applied to some simple quantum mechanical systems\cite{Gov4,Gov5}
as well as to pure U(1) Chern-Simons theory in 2+1 dimensions\cite{Gov6}, 
one of the simplest examples of a topological quantum field theory\cite{Witten}.
It would indeed be most worthwhile to explore the potential of this new
approach to the quantisation of gauge invariant systems in situations much
closer to actual theories for the fundamental interactions, either
field theories for gravity\cite{Klauder2} or the strong and electroweak
interactions, as well as any of the systems motivated by studies within
string and M-theory\cite{Schwarz}.

This contribution is organised as follows. In the next section, the discussion
of configuration space representations of the Heisenberg algebra over a
manifold of arbitrary topology and geometry is recalled. In Sect.\ref{Sect3},
Dirac's ge\-ne\-ral Hamiltonian formulation of gauge invariant systems is 
briefly described, at least in the simplest cases of such systems, concluding 
with the construction of the physical projector. Sect.\ref{Sect4} then considers
the application of the physical projector to some simple quantum mechanical
gauge invariant systems, show\-ing how the physical spectrum as well as the
wave functions of physical states
follow straighforwardly, while also circumventing any possibility
of a Gribov problem. In Sect.\ref{Sect5}, the physical projector is brought
to bear on the quantisation of the pure U(1) Chern-Simons theory in 2+1 
dimensions, leading to a physical spectrum totally independent of the
geometry of the 2+1 dimensional space but depending solely on its topology,
and in complete agreement with results achieved in the literature
through standard quantisation methods\cite{CS}.
Finally, some further remarks are presented in the Conclusions.
By its nature, the presentation of this contribution can only be sketchy;
futher details are to be found in the original 
references\cite{Gov2,Gov4,Gov5,Gov6}.

\section{Representations of the Heisenberg Algebra}
\label{Sect2}

The representation theory of the Heisenberg algebra associated to the
real line, and by extension to any flat euclidean space parametrised
by cartesian coordinates, is most familiar to anyone having studied
quantum theory. Denoting $q^\alpha$ $(\alpha,\beta=1,2,\dots,n)$ the
cartesian coordinates and $p_\alpha$ their canonical conjugate momenta,
this algebra is defined by the following commutation relations at the operator
level,
\begin{equation}
[\hat{q}^\alpha,\hat{p}_\beta]=i\hbar\delta^\alpha_\beta\ \ \ ,
\end{equation}
as well as the self-adjoint properties 
$\left(\hat{q}^\alpha\right)^\dagger=q^\alpha$
and $\hat{p}_\alpha^\dagger=\hat{p}_\alpha$. In particular, up to unitary
transformations, there exists essentially a single such representation,
corresponding to the usual plane wave representation of that algebra
(von Neumann's theorem). However, when the configuration space
coordinate system $q^\alpha$ is curvilinear, or when the configuration
space possesses a non tri\-vial geometry or even topology, usual canonical 
quantisation procedures for the associated symplectic Poisson bracket structure
through the correspondence principle---leading to the above commutation
relations defined locally over phase space---seem to have remained an open
issue. However, in the same way as may be developed for an euclidean 
space\cite{Gov1}, the representation theory of the above Heisenberg algebra
in the general case may straightforwardly be considered\cite{Gov2}.

Indeed, only two basic assumptions are required. On the one hand, that
the representation space possesses as a basis all eigenstates
$|q>$ of the position operators $\hat{q}^\alpha$ associated to
a local coordinate system set-up on the configuration space $M$, namely
$\hat{q}^\alpha|q>=q^\alpha|q>$. On the other hand, that this representation
space is equipped with an hermitian positive definite inner product for which
the operators $\hat{q}^\alpha$ and $\hat{p}_\alpha$ are indeed self-adjoint.

{}From only these two assumptions, it then follows\cite{Gov2} that the general
configuration space representation of the Heisenberg algebra is such that
the momentum operator matrix elements are given by
\begin{equation}
<q|\hat{p}_\alpha|q'>=\frac{-i\hbar}{g^{1/4}(q)}
\frac{\partial}{\partial q^\alpha}
\left(\frac{1}{g^{1/4}(q)}\delta^{(n)}(q-q')\right)+ 
\frac{1}{\sqrt{g(q)}}\,A_\alpha(q)\delta^{(n)}(q-q')\ \ ,
\end{equation}
while the vanishing commutation relations $[\hat{p}_\alpha,\hat{p}_\beta]=0$
imply the further restriction
\begin{equation}
A_{\alpha\beta}(q)\equiv\frac{\partial A_\beta(q)}{\partial q^\alpha}-
\frac{\partial A_\alpha(q)}{\partial q^\beta}=0\ \ ,\ \ 
\alpha,\beta=1,2,\dots,n\ \ .
\end{equation}
In these relations, the positive definite function $g(q)$ specifies 
the normalisation of the position eigenstates $|q>$ through
\begin{equation}
<q|q'>=\frac{1}{\sqrt{g(q)}}\,\delta^{(n)}(q-q')\ \ \ ,
\end{equation}
while the variables $A_\alpha(q)$ corresponds to a local vector field
of vanishing field strength $A_{\alpha\beta}(q)$ defined over the
configuration space.

In particular for any quantum states $|\psi>$ and $|\varphi>$, their
configuration space wave function representations $\psi(q)\equiv<q|\psi>$
and $\varphi(q)\equiv<q|\varphi>$ are such that their inner product is given by
\begin{equation}
<\psi|\varphi>=\int_Md^nq\ \sqrt{g(q)}\,\psi^*(q)\varphi(q)\ \ \ ,
\label{eq:inner}
\end{equation}
while the position and momentum operators are realised through the differential
operators,
\begin{equation}
<q|\hat{q}^\alpha|\psi>=q^\alpha\psi(q)\ \ ,\ \ 
<q|\hat{p}_\alpha|\psi>=\frac{-i\hbar}{g^{1/4}(q)}
\left[\frac{\partial}{\partial q^\alpha}+\frac{i}{\hbar}A_\alpha(q)\right]
g^{1/4}(q)\psi(q)\ \ .
\label{eq:repre}
\end{equation}

The two quantities $g(q)$ and $A_\alpha(q)$ thus parametrise all possible
representations of the Heisenberg algebra over $M$. However, not all these
representations are necessarily unitarily in\-equi\-va\-lent. Indeed, even though
$g(q)$ parametrises the normalisation of the basis states $|q>$ and
$A_\alpha(q)$ the confi\-gu\-ra\-tion space matrix elements of the
momentum operators $\hat{p}_\alpha$, we have not yet accounted for the
freedom in a possible $q$-dependent phase redefinition of the basis states,
namely $|q>_{(2)}=e^{i\chi(q)/\hbar}|q>$. In fact, such a local phase 
redefinition is tantamount to a local U(1) gauge transformation of the vector 
field $A_\alpha(q)$ which is defined through the above parametrisation of 
the $\hat{p}_\alpha$ matrix elements, from which one finds
$A^{(2)}_\alpha(q)=A_\alpha(q)+\partial\chi(q)/\partial q^\alpha$.

Hence in conclusion, the variables $A_\alpha(q)$ of vanishing field strength
determine a flat U(1) bundle over the configuration space manifold $M$,
while unitarily inequivalent representations of the Heisenberg algebra over $M$
are thus classified in terms of the gauge equivalence classes of flat 
U(1) bundles over $M$. The latter classes are characterized through the
U(1) holonomies around non contractible cycles in $M$, namely a mapping
of the first homotopy group (or fundamental group) of $M$ into the set of
gauge equivalence classes of flat U(1) connections over $M$, characterized
through their holonomies. In particular for a manifold of trivial
homotopy group, namely a simply connected one as is the case for the 
$n$ dimensional euclidean space, there is thus a single representation 
of the Heisenberg algebra, up to local U(1) unitary phase transformations in 
configuration space. This conclusion
of course corresponds to von Neumann's theorem for the real line, while
choosing then the trivial gauge configuration $A_\alpha(q)=0$ in the associated
trivial homology class provides the usual plane wave representation of that
algebra.

This conclusion still leaves open the interpretation of the normalisation
factor $g(q)$. From (\ref{eq:inner}), it is clear that through the
combination $d^nq\sqrt{g(q)}$ this function determines an integration measure
over $M$. In particular, when this integration measure is chosen to be
diffeomorphic invariant under coordinate reparametrizations in $M$,
the associated configuration space wave function representation 
$\psi(q)=<q|\psi>$ is itself diffeomorphic covariant, with the position and
momentum operators represented as in (\ref{eq:repre}). When $M$ is equipped
with a metric structure $g_{\alpha\beta}(q)$, the canonical choice for
$g(q)={\rm det}g_{\alpha\beta}(q)$ indeed determines such a diffeomorphic 
invariant integration measure, thereby leading finally to diffeomorphic
covariant representations of the Heisenberg algebra over manifolds of
arbitrary topology and geometry. An immediate example is that of curvilinear
coordinates in a flat euclidean space, but more involved cases may of course
be considered in a likewise manner.

Finally, the condition of self-adjoint momentum operators $\hat{p}_\alpha$
requires that the associated wave function representation of states be
such that
\begin{equation}
\int_Md^nq\frac{\partial}{\partial q^\alpha}\left[\sqrt{g(q)}\ 
|\psi(q)|^2\right]=0\ \ \ ,\ \ \ \alpha=1,2,\cdots,n\ \ \ ,
\end{equation}
thereby implying restrictions on states when $M$ has boundaries.

Given such a configuration space representation of the Heisenberg algebra,
it is also possible to determine the wave functions for momentum eigenstates.
However, because of the possibility of non trivial U(1) holonomies in the
general case, a network of paths $P(q_0\rightarrow q)$ connecting any point
$q$ on $M$ to a given point $q_0$ through a given path, has to be set-up
on $M$ in a continuous fashion as a function of $q$. Having done so, one
introduces the path-ordered U(1) holonomies
\begin{equation}
\Omega[P(q_0\rightarrow q)]=Pe^{-\frac{i}{\hbar}\int_{P(q_0\rightarrow q)}
dq^\alpha A_\alpha(q)}\ \ \ ,
\end{equation}
so that momentum eigenstate configuration space wave functions are given 
by\cite{Gov2}
\begin{equation}
<q|p>=\frac{e^{i\varphi(q_0,p)}}{(2\pi\hbar)^{n/2}}
\frac{\Omega[P(q_0\rightarrow q)]}{g^{1/4}(q)h^{1/4}(p)}\,
e^{\frac{i}{\hbar}(q-q_0)\cdot p}\ \ \ ,
\end{equation}
where $\varphi(q_0,p)$ is an arbitrary phase factor, while the function
$h(p)$ determines the normalisation of the momentum eigenstates according to
\begin{equation}
<p|p'>=\frac{1}{\sqrt{h(p)}}\delta^{(n)}(p-p')\ \ \ .
\end{equation}
Clearly, in the case of the trivial representation with $A_\alpha(q)=0$,
one recovers the usual plane wave representation of the Heisenberg algebra,
extended to include the normalisation factors $g(q)$ and $h(p)$ possibly
different from unity in order to ensure proper diffeomorphic covariant
properties in the case of non cartesian coordinates over $M$.

Given these different expressions, it then also becomes possible to
set-up a phase space path integral representation of matrix elements of quantum
operators, say in configuration space. The ensuing expressions are
identical to the usual ones in discretized form, the sole difference
appearing through the normalisation and the path dependency of the external
states (see the original reference\cite{Gov2} for further details).

Once such a general discussion of representations of the Heisenberg algebra
over configuration spaces of arbitrary topology and geometry has been
displayed, it is a simple matter to apply it to any given system whose dynamics
is determined through some action principle. Based on the latter, the canonical
Hamiltonian formulation of the system may be developed, which through
canonical quantisation then leads to a certain quantum representation
of the quantised system. In the case of a configuration space of non trivial
first homotopy group, the choice of representation parametrised in terms of
non trivial U(1) holonomies of the flat bundle around the non contractible
cycles in $M$ is a matter of physics, in the same way that for example the spin 
representation of a quantised system invariant under space rotations is a matter
of physics. It is also at the level of the Hamiltonian that a possible
metric structure over $M$ may appear and thus determine the adequate
reparametrisation covariant representation of the Heisenberg algebra
to be used. Finally, in correspondence with the given classical Hamiltonian,
a two parameter class of hermitian (and possibly self-adjoint)
quantum Hamiltonian operators may be introduced\cite{Gov2}, which all reduce 
to the classical one in the limit $\hbar\rightarrow 0$. Again, further
developments are left for the interested reader to find in the original
reference. The main point this contribution wished to emphasize is the general
classification of configuration space representations of the Heisenberg
algebra over a manifold of arbitrary topology and geometry in terms of
topology and gauge equivalence classes of flat U(1) bundles over that
manifold.

\section{Gauge Invariant Systems and the Physical Projector}
\label{Sect3}

The analysis of the Hamiltonian formulation of constrained
systems goes back to Dirac\cite{Gov1} and is very general. Here, only the
simplest possible situation will be outlined explicitely, corresponding
essentially to systems with gauge symmetries of the Yang-Mills type.
Moreover, we shall also assume that all degrees of freedom are Grassmann
even variables which thus commute at the classical level.

As is well known, when a system possesses continuous (global) symmetries,
by Noether's theorem there exist conserved quantities generating these
symmetries, whose Poisson brackets among one another also determine 
the Lie algebra of the symmetry group, while their Poisson brackets with the 
Hamiltonian of the system close among themselves. More explicitely, 
if $H(q,p)$ denotes the latter Hamiltonian while $\phi_a(q,p)$ stand for the 
symmetry generators, one has
\begin{equation}
\{\phi_a(q,p),\phi_b(q,p)\}={C_{ab}}^c\phi_c(q,p)\ \ ,\ \ 
\{H(q,p),\phi_a(q,p)\}={C_a}^b\phi_b(q,p)\ \ \ .
\end{equation}
Here, ${C_{ab}}^c$ and ${C_a}^b$ are specific coefficients assumed
to be constants in our discussion (more generally however, these coefficients 
may be phase space dependent quantities\cite{Gov1}).

In the case of local or gauge symmetries, these properties of the 
conserved quantities are of course preserved, but in addition, the physical 
requirement of gauge invariance implies further that 
these conserved quantities must vanish identically, $\phi_a(q,p)=0$,
for gauge invariant configurations. Moreover, the dynamics
of the system in phase space is then governed by the action principle
(assuming $(q^\alpha,p_\alpha)$ to define canonically conjugate degrees
of freedom\cite{Gov1})
\begin{equation}
S[q,p;\lambda^a]=\int_{t_i}^{t_f}dt\left[\dot{q}^\alpha p_\alpha
-H(q,p)-\lambda^a\phi_a(q,p)\right]\ \ \ ,
\end{equation}
where $\lambda^a(t)$ are arbitrary Lagrange multipliers for the gauge generator
constraints $\phi_a=0$, which also parametrise the gauge freedom generated
by $\phi_a$ under time evolution. In particular, infinitesimal gauge
transformations then correspond to the variations
\begin{equation}
\delta_\epsilon q^\alpha=\{q^\alpha,\epsilon^a\phi_a\}\ ,\ 
\delta_\epsilon p_\alpha=\{p_\alpha,\epsilon^a\phi_a\}\ ,\ 
\delta_\epsilon \lambda^a=\dot{\epsilon}^a+
\lambda^c\epsilon^b{C_{bc}}^a-\epsilon^b{C_b}^a\ \ ,
\end{equation}
where $\epsilon^a(t)$ are arbitrary infinitesimal functions.

Within Dirac's quantisation of such systems\cite{Gov1}, the usual rules
of canonical quantisation are simply applied to all the phase space
degrees of freedom $(q^\alpha,p_\alpha)$, leading {\em a priori\/}
to representation algebras of the Heisenberg type as discussed in the
previous section. The restrictions of the gauge invariance are then
imposed through the operator constraints
\begin{equation}
\hat{\phi}_a|\psi>=0\ \ \ ,
\end{equation}
while time evolution is generated by the time ordered operator
\begin{equation}
S(t_f,t_i)=Te^{-\frac{i}{\hbar}\int_{t_i}^{t_f}dt
\left[\hat{H}+\lambda^a(t)\hat{\phi}_a\right]}\ \ \ .
\end{equation}

Clearly, such an operator propagates both gauge invariant as well as
gauge variant states, and thus cannot correspond to the physical
propagator of the system to which only physical, {\em i.e.\/} gauge invariant
states would contribute. Usually, in order to achieve that aim,
one considers some gauge fixing procedure through which only gauge invariant
configurations are maintained in the actual time dependent dynamics.
This gauge fixing may be effected before quantisation, namely through the
so-called Faddeev reduced phase space approach\cite{Gov1}. Or else, the
gauge fixing is effected through the so-called BRST quantisation of the
gauge invariant system\cite{Gov1}, in which case the original phase space
is extended to include degrees of freedom---namely ghosts---of Grassmann parity
opposite to that of the original degrees of freedom, in order to compensate
for the contributions of the gauge variant configurations and of
the degrees of freedom conjugate to the Lagrange multipliers $\lambda^a$ 
which are also introduced in order to render the $\lambda^a$'s
dynamical as well. However, whatever the approach to gauge fixing being
implemented, even though the ensuing description is by construction indeed
always gauge invariant, it may ge\-ne\-ral\-ly suffer Gribov problems\cite{Gov1},
namely it may include some gauge invariant configurations more than once,
or not at all as the case may be, clearly an unacceptable situation if
one is to properly include once and only once all physically distinct
configurations possibly accessible to the system.

Nevertheless, Gribov problems are a consequence of gauge fixing, so that if the
latter may be circumvented, Gribov problems would simply not be an issue
anymore. The recently proposed physical projector approach to the quantisation
of constrained systems\cite{Klauder1} indeed provides such a framework
which avoids the apparent necessity of gauge fixing, while at the same
ensuring a proper inclusion of all physically distinct configurations 
and thus avoiding Gribov problems altogether\cite{Gov3}. In fact, the
physical projector approach is simply set within Dirac's formulation
briefly described above, and uses the projector onto the gauge invariant
components of any quantum state of the quantised system. The projector
itself is obtained by integrating over the group of gauge transformations
the transformations of states generated by the gauge generators $\hat{\phi}_a$,
\begin{equation}
{\cal E}=\int dU(\theta^a)e^{-\frac{i}{\hbar}\theta^a\hat{\phi}_a}\ \ \ .
\end{equation}
Here, $\theta^a$ are coordinates over the group manifold, while $dU(\theta^a)$
is the group invariant integration measure normalised such that ${\cal E}$
indeed obeys the pro\-per\-ties of a projection operator,
\begin{equation}
{\cal E}^2={\cal E}\ \ \ ,\ \ \ {\cal E}^\dagger={\cal E}\ \ \ .
\end{equation}
In the case of a compact gauge group, this definition of ${\cal E}$
is sufficient. When the gauge group is non compact, thereby leading to
a spectrum of the gauge generators $\hat{\phi}_a$ which is not purely
discrete, further considerations need to be applied\cite{Klauder1}.

Given the physical projector ${\cal E}$, the definition of the physical
evolution operator propagating gauge invariant states only is clearly,
\begin{equation}
S_{\rm phys}(t_f,t_i)={\cal E}\,S(t_f,t_i)\,{\cal E}=
e^{-\frac{i}{\hbar}(t_f-t_i)\hat{H}}\,{\cal E}=
{\cal E}\,e^{-\frac{i}{\hbar}{\cal E}\hat{H}{\cal E}}\,{\cal E}\ \ \ ,
\end{equation}
making it explicit that indeed only gauge invariant states contribute to
this operator both as external states as well as intermediate ones.
In particular, in the case of a compact Lie group of gauge transformations,
as implicitely assumed in this discussion having taken the structure
coefficients ${C_{ab}}^c$ to be constants, it is clear that the evaluation
of matrix elements of such a physical operator, and of the physical
projector itself, immediately implies group theory techniques for the
calculation of various group invariants.

In the remainder of this contribution, different examples of the application
of the physical projector quantisation of gauge invariant systems are
briefly described.

\section{Some Quantum Mechanical Gauge Invariant Systems}
\label{Sect4}

One of the simplest examples of quantum mechanical systems possessing
a local gauge invariance is that of a SO(2)=U(1) local gauge invariance
in a plane defining two degrees of freedom. However, this situation being
far too simple by itself, the invariance under time dependent SO(2) rotations
may be coupled to translations in a direction perpendicular to the
plane of rotations, thereby leading finally to a system with three
degrees of freedom of cartesian coordinates $(x(t),y(t),z(t))$, whose
dynamics is governed by the Lagrange function\cite{Lee},
\begin{equation}
L=\frac{1}{2}\left[(\dot{x}+g\xi y)^2+(\dot{y}-g\xi x)^2+
(\dot{z}-\xi)^2\right]-V(\sqrt{x^2+y^2})\ \ \ .
\end{equation}
Here, $\xi(t)$ is a gauge degree of freedom, $g$ is a gauge
coupling constant, while the potential $V(\sqrt{x^2+y^2})$ is
a rotation invariant quantity which for later purposes is taken to be
a spherically symmetric harmonic potential $V=\omega^2(x^2+y^2)/2$.

Since gauge transformations in this system correspond to time dependent
SO(2) rotations in $(x(t),y(t))$ coupled to time dependent translations
in $z(t)$, it should be clear that the generator of this local abelian
gauge symmetry is given by
\begin{equation}
\phi=p_z+gp_\theta\ \ \ ,
\end{equation}
where $p_z$ is the momentum conjugate to the coordinate $z$ while $p_\theta$
is that conjugate to the angular variable $\theta$ associated to polar
coordinates $(r,\theta)$ in the $(x,y)$ plane. Similarly, the gauge invariant 
Hamiltonian of the system simply reads,
\begin{equation}
H=\frac{1}{2}p^2_r+\frac{1}{2r^2}p^2_\theta+\frac{1}{2}p^2_z+
\frac{1}{2}\omega^2r^2\ \ \ ,
\end{equation}
which clearly has a vanishing Poisson bracket with the gauge generator $\phi$,
since the variables $(r,p_r)$, $(\theta,p_\theta)$ and $(z,p_z)$ each 
form a pair of canonically conjugate phase space degrees of freedom.

Canonical quantisation of the system is straighforward enough\cite{Lee,Gov5},
and is best developed in the helicity basis\cite{Gov4} related to the
SO(2) symmetry in the $(x,y)$ plane. Given the creation and annihilation
operators $a^{(\dagger)}_{x,y}$ associated to the cartesian coordinates
$(x,y)$ in the usual way, the helicity creation and annihilation operators
are defined by
\begin{equation}
a_{\pm}=\frac{1}{\sqrt{2}}\left[a_x\mp ia_y\right]\ \ \ ,\ \ \ 
a^\dagger_{\pm}=\frac{1}{\sqrt{2}}\left[a^\dagger_x\pm ia^\dagger_y\right]
\ \ \ .
\end{equation}
One then finds
\begin{equation}
\hat{H}=\frac{1}{2}\hat{p}^2_z+\hbar\omega\left[
a^\dagger_+a_++a^\dagger_-a_-\right]\ \ \ ,\ \ \ 
\hat{\phi}=\hat{p}_z+\hbar g\left[
a^\dagger_+a_+-a^\dagger_-a_-\right]\ \ \ .
\end{equation}
These expressions make obvious what the physical spectrum of the system is,
and how to construct the associated quantum excitations of the Fock vacuum.
In particular, working in the configuration space representation of the
Heisenberg algebra associated to the polar coordinates $(r,\theta,z)$,
it is then straightforward to determine the wave functions of all gauge
invariant states\cite{Gov5}. For example, the physical energy spectrum is
thus given by
\begin{equation}
E_{n_{\pm}}=\frac{1}{2}p^2+\hbar\omega(n_++n_-+1)\ \ \ ,\ \ \ 
{\rm with}\ \ \ p=-\hbar g(n_+-n_-)\ \ \ ,
\end{equation}
$n_\pm$ being of course the integer excitation numbers of the modes
of left- or right-handed SO(2) helicity.

For the present system, the spectrum of the gauge generator $\hat{\phi}$
being continuous (because of the $\hat{p}_z$ contribution), the definition
of the physical projector requires first to consider the projector 
${\cal E}_\delta$ onto those states whose $\hat{\phi}$ eigenvalue lies
within the interval $[-\delta,+\delta]$, $\delta>0$ being a positive
number taken to be as small as may be required. The operator ${\cal E}_0$
projecting onto those states such that $\hat{\phi}=0$ is then obtained
as\cite{Klauder1,Gov5,Gov6}
\begin{equation}
{\cal E}_0=\lim_{\delta\rightarrow 0}\frac{\pi\hbar}{\delta}{\cal E}_\delta=
\int_{-\infty}^{+\infty}d\gamma\,e^{\frac{i}{\hbar}\gamma\hat{\phi}}\ \ \ .
\end{equation}

Configuration space matrix elements of the physical propagator
of the system, $S_{\rm phys}(t_f,t_i)=e^{-i(t_f-t_i)\hat{H}/\hbar}{\cal E}_0$,
are then readily determined. An explicit calculation\cite{Gov5} finds,
\begin{equation}
\begin{array}{r l}
<r_f,\theta_f,z_f|&S_{\rm phys}(t_f,t_i)|r_i,\theta_i,z_i>=
\frac{\omega}{2i\pi\hbar\sin\omega\Delta t}
e^{\frac{i\omega\cos\omega\Delta t}{2\hbar\sin\omega\Delta t}(r^2_f+r^2_i)}
\times\\
 & \\
\times&\sum_{\ell=-\infty}^{+\infty}\,e^{-i\frac{\pi}{2}|\ell|}
e^{i\ell(\varphi_f-\varphi_i)}e^{-\frac{i}{2}\hbar\Delta tg^2\ell^2}
J_{|\ell|}\left(\frac{\omega r_fr_i}{\hbar\sin\omega\Delta t}\right)\ \ \ ,
\end{array}
\end{equation}
where $\Delta t=t_f-t_i$ and $\varphi_{f,i}=\theta_{f,i}-gz_{f,i}$.
In fact, given this final expression, it is possible\cite{Gov5} to show
that each single physical state of the system contributes to these matrix
elements once and only once as an intermediate state, thereby establishing
that the physical projector approach is indeed necessarily free of any
Gribov problem. Moreover, it is also possible from this expression of the
physical propagator to extract the wave functions for each of the physical
states, the above matrix elements playing somehow the r\^ole of generating
functions for these wave functions.

Other similar examples of gauge invariant quantum mechanical systems may
be considered. For instance, rather than introducing the $z(t)$ degree
of freedom associated to the coupling of the time dependent local SO(2)
gauge transformations with time dependent translations in the $z$ direction,
one may consider a whole collection of $N$ similar $(x_i(t),y_i(t))$ 
SO(2) invariant pairs of degrees of freedom $(i=1,2,\cdots,N)$, which are all 
rotated by the same time dependent angle irrespective of the value of the 
index $i$. SO(2) gauge invariance then requires that the total SO(2) angular 
momentum of such systems be vanishing, thereby leading to a compact group 
of gauge transformations. Here again when introducing an identical SO(2) 
invariant harmonic potential for all these degrees of freedom, the quantisation
of such systems is straightforward, while the physical projector approach
readily leads to the determination of the physical spectrum and of the
wave functions of its physical states, whether in configuration space or
any other choice of basis for quantum states, such as coherent 
states\cite{Gov4}. 

Such considerations have been worked out in all details
for the SO(2) case, and to some extent for the SO(3) case\cite{Gov3}. 
Generalisations to all Lie algebras are possible\cite{Gov3}, and would be 
worth exploring. In particular, it would be extremely interesting to combine
techniques of coherent states with those of combinatorics in group theory,
which is possible now through the use of the physical projector. Indeed,
as the above construction of the latter ope\-ra\-tor should have made clear, 
matrix elements of the physical projector are given by integrals over the group 
manifold of the exponentiated generators, hence leading to specific combinations
of character invariants in the group. Unravelling the connection between
these characters and the quantities to be integrated over the group manifold
may provide new combinatorial identities for the representation theory of
compact Lie algebras. Exceptional Lie algebras may be particularly
fascinating cases in that respect, beginning with the case of $G_2$.

\section{U(1) Chern-Simons Theory and the Physical Projector}
\label{Sect5}

Given the space available, let us now turn to an example of a quantum
field theory, rather than a quantum mechanical system, whose gauge freedom
is so large that gauge invariance projects down a finite number of physical
con\-fi\-gu\-ra\-tions only from the initially infinite number of degrees of 
freedom.  Such theories are dubbed topological quantum field theories, since 
their sets of physical states only depend on the topology of the underlying
manifold on which they are constructed\cite{Witten}. As a matter of fact,
the physical projector approach is also apt to handle the intricacies
of such systems\cite{Gov6}, as briefly described in this section.

Among the simplest of such topological quantum field theories are the
U(1) Chern-Simons theories in 2+1 dimensions, whose action principle reads
\begin{equation}
S=N_k\int_Mdx^0dx^1dx^2\epsilon^{\mu\nu\rho} A_\mu\partial_\nu A_\rho\ \ \ .
\end{equation}
Here $N_k$ is a normalisation factor taking quantised values at
the quantum level\cite{Witten} (hence the notation), while $A_\mu(x^\mu)$ 
$(\mu,\nu=0,1,2)$ is a U(1) gauge connection defined over the 2+1 dimensional 
base manifold $M$.  That this system is topological in character is obvious 
from two facts, The first is that the above action is defined irrespective of 
a metric structure on $M$; only a differentiable structure is required. 
The second fact is that the equations of motion of the system reduce to
\begin{equation}
F_{\mu\nu}=\partial_\mu A_\nu-\partial_\nu A_\mu=0\ \ \ .
\end{equation}
Hence, the dynamics of the system is that of flat U(1) bundles over $M$,
whose characterization is provided entirely through the U(1)
holonomies of the gauge field $A_\mu(x^\mu)$ in $M$, which indeed is
purely a topological issue determined by the topology of $M$.

For the purpose of canonical quantisation, let us restrict to topologies
of the form $M=\Sigma\times I\!\!R$, where $\Sigma$ is an arbitrary
compact Riemann surface, later on to be taken to be a torus $T$ (note that
the 2+1 dimensional split does not refer to a metric signature, but to such
a topology split in the diffeomorphic structure of $M$). In such a case,
the holonomies of $A_\mu$ reduce to the holonomy classes around the
cycles defining a basis of the first homology group of $\Sigma$.
These holonomies only involve the zero mode contributions of the gauge field,
so that the entire physical content of the system is restricted to lie
entirely within its zero mode sector. Hence in order to present results,
let us only concentrate on that sector, even though the whole canonical
quantisation procedure may be applied to all degrees of freedom,
while local U(1) gauge transformations homotopic to the
trivial one ({\em i.e.\/} ``small" gauge transformations) may be 
used\cite{Gov6} to gauge away any non zero mode configuration for the 
field $A_\mu(x^\mu)$.

Specifically, let us consider the case of the torus $T$, having only
two homology cycles. The coordinate system associated to these cycles
enables a mode decomposition\cite{Gov6} of the field $A_\mu(x^\mu)$. Under
large gauge transformations characterized by the two integer holonomies
$k_1$ and $k_2$ associated to these two cycles, the zero modes $A_1$ 
and $A_2$ transform according to
\begin{equation}
A'_1=A_1+2\pi k_1\ \ \ ,\ \ \ A'_2=A_2+2\pi k_2\ \ \ .
\label{eq:large}
\end{equation}
On the other hand, these two degrees of freedom define in fact the phase
space of the system, with in particular the commutation relations
\begin{equation}
[\hat{A}_1,\hat{A}_2]=i\frac{\hbar}{2N_k}\ \ \ .
\end{equation}
Hence this system is distinguished by having a phase space which itself
is a compact 2-torus of volume $(2\pi)^2$, quite a unique feature not
directly amenable to usual quantisation techniques. Moreover, the above
commutation relation shows that the number of physical states must also
be given by $(2\pi)^2/(2\pi\hbar/(2N_k))=4\pi N_k/\hbar$, 
a first indication that the factor $N_k$
indeed needs to be quantised at the quantum level ($N_k>0$ is
assumed; when $N_k$ is negative,
the r\^oles of the coordinates $x^1$ and $x^2$ are interchanged).

A coherent state quantisation of the system requires some further
structure to be introduced beyond the mere topological and diffeomorphic ones,
namely a complex structure parametrised by a complex parameter
$\tau=\tau_1+i\tau_2$ such that $\tau_2>0$. The associated annihilation 
and creation operators are then defined by
\begin{equation}
\alpha=\sqrt{\frac{N_k}{\hbar\tau_2}}\left[-i\tau\hat{A}_1+i\hat{A}_2\right]
\ \ ,\ \ 
\alpha^\dagger=\sqrt{\frac{N_k}{\hbar\tau_2}}
\left[i\bar{\tau}\hat{A}_1-i\hat{A}_2\right]\ \ .
\end{equation}
Having restricted our considerations to the zero mode sector alone,
the construction of the physical projector relates to large
gauge transformations ac\-ting on the zero modes only. From the gauge
transformations in (\ref{eq:large}), this projector is found\cite{Gov6} to be
given by ${\cal E}=\sum_{k_1,k_2=-\infty}^{+\infty}\hat{U}(k_1,k_2)$ with
\begin{equation}
\hat{U}(k_1,k_2)=C(k_1,k_2){\rm exp}\left\{\frac{2i\pi}{\hbar}
\sqrt{\frac{\hbar N_k}{\tau_2}}\left[
(\bar{\tau}k_1-k_2)\alpha+(\tau k_1-k_2)\alpha^\dagger\right]\right\}\ \ \ .
\end{equation}
Here, $C(k_1,k_2)$ is a cocyle factor such that the group composition
law is obeyed for the above operator representation of large gauge
transformations. The latter requirement implies that\cite{Gov6}
\begin{equation}
C(k_1,k_2)=e^{i\pi k k_1 k_2}\ \ \ ,\ \ \ {\rm with}\ \ \ 
N_k=\frac{\hbar}{4\pi}k\ \ \ ,
\end{equation}
where $k$ is an arbitrary strictly positive integer, hence also to be equal to
the number of physical states.

Within the coherent state representation of the system, it is now possible
to determine its physical spectrum content as well as the coherent state
wave functions for these gauge invariant states, from the simple application
of the physical projector onto the space of quantum states. 
An explicit analysis\cite{Gov6} finds that the number of physical states 
is indeed equal to the integer $k$ which quantises the normalisation 
factor $N_k$, while the obtained wave functions agree completely
with results established previously in the literature\cite{CS}
(see the ori\-gi\-nal references\cite{Gov6,CS} for explicit expressions).
By construction, these physical states $|r>$ $(r=0,1,\cdots,k-1)$ are
gauge invariant under large gauge transformations. However, to which
extent a dependency on the complex structure $\tau$ of the torus $T$ arises
also needs to be assessed. The explicit analysis\cite{CS,Gov6} of this
issue finds that this further requirement of modular invariance in the
parameter $\tau$ implies that the integer $k$ must also be even,
while the whole set of physical states then provides a single
irreducible representation of the mo\-du\-lar group of the Riemann surface
$\Sigma$. This is thus how
the topological invariance of the original classical theory
is characterized at the quantum level. The dependency on the complex
structure through modular classes is a consequence of a conformal
anomaly of the quantised system\cite{Witten}.

\section{Conclusions}
\label{Sect6}

As mentioned in the Introduction, the purpose of this contribution is
to highlight some recent results concerned with the issues raised by the
canonical quantisation of systems whose configuration space is not
euclidean and flat or is parametrized by curvilinear coordinates,
and of systems pos\-ses\-sing local gauge invariances. A general classification
of representations of the Heisenberg algebra in the first case was briefly
described, which implies the appearance of topology classes of
flat U(1) bundles over configuration space, together with a diffeomophic
covariant wave function representation of quantum states. With respect
to gauge invariant systems, a new approach\cite{Klauder1} to their quantisation
has been discussed, and shown through explicit examples to lead to a perfectly
adequate framework avoiding any gauge fixing and thereby also the possibility
of Gribov problems. Other advantages of the physical projector approach
have not been emphasized however, such as for example the fact that it is
also apt to address the issues raised by the quantisation of compact phase
spaces with isometries, as illustrated through the U(1) Chern-Simons
theory example.

The next stage of explorations based on the physical projector
is now to be initiated, aiming towards a deeper insight into the
non perturbative dynamics of theories ever closer to those of the
actual material universe and its quantum excitations, whether within
the realm of quantum field theories or other inceptions instigated by
the recent developments in string and M-theory.

\section*{Acknowledgements}

Prof. J.R. Klauder is gratefully acknowledged for many useful discussions
over the years, and for his constant interest in this work. 
Prof. M.N. Hounkonnou is thanked for his insistence in
getting this contribution into the written form, and more importantly
for the initiative and the organization of this Workshop,
which hopefully should be the first in a long series of such meetings
of ever increasing purposefulness and interest.

\end{document}